# Direct Detection and Spectral Characterization of Outer Exoplanets with the SPICA Coronagraph Instrument (SCI)

Taro Matsuo*,[1], Misato Fukagawa[2], Takayuki Kotani[3],
Yoichi Itoh[4], Motohide Tamura[1], Takao Nakagawa[3],
Keigo Enya[3], & SCI team

1) National Astronomical Observatory of Japan, 2-21-1, Osawa, Mitaka, Tokyo 181-0015, Japan
2) Department of Earth and Space Science, Osaka University, 1-1, Machikaneyama, Toyonaka, Osaka 560-0043, Japan
3) Department of Infrared Astrophysics, Institute of Space and Astronautical Science, Japan Aerospace Exploration Agency, 3-1-1, Yoshinodai, Chuo-ku, Sagamihara, Kanagawa 252-5210, Japan
4) Graduate School of Science, Kobe University, 1-1, Rokkodai, Kobe 657-8501, Japan

Email address: taro.matsuo@nao.ac.jp

## Abstract

The SPICA coronagraph instrument (SCI) provides high-contrast imaging and moderate resolution (R<200) spectroscopy at the wavelength range from 3.5 to 27 µm. Based on the planet evolutional model calculated by Burrows et al. (2003), SCI will search for gas giant planets down to one Jupiter mass around nearby young (1 Gyr) stars and two Jupiter masses around nearby old (5 Gyr) stars. SCI also allows to characterizing those planets of less than 1 Gyr by spectroscopic observations to reveal the nature of planetary formation and evolution.

Focusing on the high sensitivity and high contrast at wavelengths longer than 10 µm, we show that SCI also allows us to directly image icy giant planets like Uranus and Neptune as well as gas giant planets around nearby early-type stars. In this paper, we compare the capabilities of SCI and the JWST coronagraphs and also discuss a new approach to answering questions concerning the formation and evolution of planetary systems through planet detection with SCI.

## 1. Planet detection with the SPICA coronagraph
## 1.1 Introduction

Since the first discovery of an exoplanet around a normal star by Mayor & Queloz (1995), more than 450 exoplanets have so far been discovered. This success has largely been due to the accurate radial velocity (RV) method, with which more than 80 % of the known exoplanets as well as their surprising dynamical diversity have been discovered. However, these discoveries are inherently limited in several ways. (1) Since the radial velocity studies are confined to the inner regions of extrasolar planetary systems (<6AU for a 15-yr survey), we still know very little about the planetary constituents in the outer regions. (2) RV surveys of "young" stars are complicated due to the high level of intrinsic stellar activity; RV surveys have thus traditionally targeted old and quiet stars. In addition, the RV method allows us to gain knowledge of only the physical properties such as semi-major axis, minimum planet-mass, eccentricity, and radius.

On the other hand, direct imaging is one of the best ways to investigate the surroundings of young stars, especially the "initial" distribution of the planets, and direct spectroscopy can tell us about the chemical compositions of their atmospheres. The direct method is very important not only for characterization of exoplanets but also for study on planet formation. Chauvin et al. (2004) reported the first successful direct detection of a planetary mass companion orbiting a brown dwarf with VLT. High contrast imaging observations have since been performed with large ground-based telescopes such as Subaru, Gemini, and VLT. Recently, Marois et al. (2008) and Kalas et al. (2008) successfully imaged exoplanets orbiting the A-type Vega-like stars of HR 8799 and Fomalhaut with Keck/Gemini and HST, respectively. Very recently, Lagrange et al. (2010) also confirmed a planet around beta Pic with VLT, originally reported in 2008. Regarding planets around G stars, Subaru has discovered a potentially planetary-mass companion around the G9 star GJ 758 (Thalmann et al. 2009). The latter is the first detection of a fairly old (700 My or more) system. However, it is in general difficult to directly detect planets around old planetary systems due to the limited contrast of the near infrared coronagraphs mounted on these ground-based telescopes.

We note that Burrows (2003), Baraffe et al. (2003), and Fortney et al. (2008) numerically calculated the evolutionary tracks for brown dwarfs and planets. According to these models, for old planetary systems, the contrast ratios of planets to stars in the mid-infrared region are much smaller than those in the near infrared. This is because young planets are still bright in the mid-infrared but dramatically cool down as they age. Therefore, direct imaging of cool planets in the mid-infrared is imperative for studying the formation and evolution of planetary systems through observations of both young and old planetary systems.

1.2. SPICA coronagraph

The Space Infrared telescope for Cosmology and Astrophysics (SPICA) is a proposed space mission for mainly mid-to-far infrared (MIR/FIR) astronomy, led by the Japan Aerospace Exploration Agency (JAXA) (Nakagawa et al. 2010), and is planned to be launched in FY2018. The SPICA carries a single 3.2-m aperture space telescope cooled to 6K with focal plane instruments and provides a substantial sensitivity advantage in the infrared wavelength region. It will allow us to tackle a wide range of open questions from the formation and evolution of stars and galaxies in the early universe to the origin and evolution of planetary systems.

One of the proposed instruments for the SPICA is a mid-infrared coronagraph for studying exoplanets, called the SPICA Coronagraph Instrument (hereafter SCI, P.I.: K. Enya, Enya et al. 2010 and its references, see also Tamura 2000). SCI provides a high contrast of $10^{-6}$ at the inner working angle (IWA), and potentially $\sim 10^{-7}$ after PSF subtraction, over a wide observing wavelength range from 3.5 to 27 μm thanks to an achromatic binary shaped pupil mask, which has been selected as the current baseline of SCI in terms of the feasibility.

On the other hand, the James Webb Space Telescope (JWST) is a space mission, which has a 6.5m segmented aperture with four types of instruments, and will be launched in 2014. Several coronagraphs will be mounted on JWST: the NIRCam coronagraph between 2.1 μm and 4.6 μm (Krist et al. 2010), the Non-Redundant Masking on the Fine Guidance Sensor Tunable Filter Imager (FGS-TFI/NRM) between 3.8 μm and 5.5 μm

(Sivaramakrishnan et al. 2009), and the Four-Quadrant Phase Mask on the Mid-Infrared Instrument (MIRI/FQPM) for mid-infrared wavelengths longer than 10 μm (Cavarroc et al. 2008). Relative to the JWST coronagraphs, SCI will have significant advantages in contrast and sensitivity at the observing wavelengths longer than 10 μm; however the IWA of SCI is limited. SCI also has a high sensitivity so that even planetary-mass objects can be detected. Therefore, SCI gives very promising opportunities for the direct detection and spectroscopy of very cool gas giants in the outer regions around old systems. Furthermore, focusing on the fact that objects irradiated by the early-type stars are still warm even if their semi-major axes are beyond 20 AU, we show that SCI can directly detect outer icy giants orbiting early-type stars from high contrast observations at wavelengths longer than 10 μm. In this paper, we propose our new approach to answering questions concerning the formation and evolution of planetary systems through planet detection with SCI, comparing its expected performance with that of JWST.

## 2. Expected performance of SCI
## 2.1. Direct detection

In this section, we show the expected limiting performances for planet detection with SCI and the JWST coronagraphs, and then show how SCI is a unique instrument for exoplanet studies. First, we summarize the important parameters of SCI and the JWST coronagraphs. Table 1 is a compilation of the IWA, the outer working angle (OWA), the achievable contrast, the sensitivity, the observing wavelengths, and the spectral resolution of SCI and the JWST coronagraphs. Here, according to Beichman et al. (2009), JWST uses FGS-TFI/NRM at 4.44 μm, the spot coronagraph on NIRCam at 4.44 μm, and MIRI/FQPM at 11.4 μm for the planet survey. Based on these parameters and "Hot start" models, the COND03 model (Baraffe et al. 2003) and the TLUSTY model (Burrows et al. 2003), we evaluated the detection limits of planets around young (1 Gyr) and old (5 Gyr) nearby G stars at 10 pc for SCI and the JWST coronagraphs. The reason why the two evolutional models are used is that the difference between the two models is very large. Note that the limiting performances are not derived based on the

"Core-accretion" model presented by Fortney et al. (2008) because the TLUSTY and the "Core-accretion" models agree well with each other after 1 Gyr (see Marley et al. 2007).

Figure 1 shows the limiting performances of SCI and the coronagraphs used for the JWST planet survey, based on the COND03 model for very cool objects down to 100 K, at which point the model calculation starts to become unreliable (Baraffe et al. 2003). FGS-TFI/NRM can search for massive gas giants near the young stars thanks to the very small IWA while the limited IWAs of NIRCam/spot, MIRI/FQPM, and SCI forces you to look at far-out planets since you can't see the inner ones. In order to see low-mass planets at large separations, a high sensitivity is needed instead. As in the case of the young planetary systems, the JWST coronagraphs can search for gas giants close to and far from their host stars while SCI can search for very low-mass objects down to two Jupiter masses in the outer regions (>10 AU). Thus, compared with SCI, the combination of the JWST coronagraphs has an advantage in planet detection around both young and old nearby stars.

Next, we evaluate the limiting performances of SCI and the JWST coronagraphs for planet survey using the TLUSTY model for T dwarfs and planetary-mass objects with an effective temperature range from 800 K to 150 K (see Figure 2). Compared with the COND03 model, a large dynamic range is required for direct detection of the planetary-mass objects. As a result, only SCI, which has a much higher contrast than the JWST coronagraphs, can reach very low mass objects down to one Jupiter mass for the young systems and two Jupiter masses for the old systems. On the other hand, the FGS-TFI/NRM can only search for brown dwarfs around the old G stars due to the smaller dynamic range and the NIRCam coronagraph can reach planetary mass objects only far from the host stars. Therefore, SCI is unique in its capability to search for planetary-mass objects in old planetary systems. In addition to this, SCI can also explore an adequately wide orbital range around a star with the same detection limit. Thus, regarding the limiting performances derived through the TLUSTY model, SCI gives us a unique opportunity for studying the exoplanets through observations of both young and old planetary systems.

|  | Wavelength (μm) | IWA (arcsec) | OWA (arcsec) | 5-sigma dynamic range after subtraction (mag) | Sensitivity limit (5σ, 1 h mag) | Spectral resolution |
|---|---|---|---|---|---|---|
| SPICA coronagraph instrument (SCI) | 3.5–27 | 2.19 at 10 μm (3.5 l/D) | 10.0 at 10 μm 20.0 at 20 μm | 17.5 | 17.8 at 12 μm 15.8 at 25 μm[c] | 5, 20, 200 |
| JWST FGS-TFI/NRM (Sivaramakrishnan et al. 2009) | 3.8–5.5 | 0.07 at 4.6 μm (0.5 l/D) | 0.55 (4 l/D) | 12.5 w. Cal. 10.0 w/o Cal.[a] | 20.6[a] | 5, 100 |
| JWST NIRCam/occulting spot mask (Beichman et al. (2010)) | 2.1, 3.35, and 4.3 | 0.56 at 4.3 μm (6 l/D) | 10.0 | For the 4.3 μm mask, 9.9 at 0″.5 11.7 at 1″ 14.3 at 2″ 16.2 at 4″[a] | 24.8 at 3.35 μm 23.6 at 4.4 μm[a] | 4, 10 |
| JWST NIRCam/occulting wedge-shaped mask (Beichman et al. (2010)) | 2.1 and 4.6 | 0.85 at 4.6 μm (4 l/D) | 10.0 | Same as the occulting spot[b] | - Same as the occulting spot | 4, 10 |
| JWST MIRT/FQRM (Beichman et al. (2010)) | 10.65, 11.4, and 15.5 | 0.36 at 11.4 μm (1 l/D) | 13.0 | For the 11.4 μm mask, 9.0 at 0″.5 9.5 at 1″ 12 at 2″ 13 at 4″[a] | 17.6 at 11.4 μm[a] | 20 |

[a] The dynamic ranges and the sensitivity limits of the JWST coronagraphs are based on Table 4 in Beichman et al. (2010).
[b] The spot occulter performs slightly better at larger separation than the wedge-shaped one.
[c] SCI's sensitivity limit includes a degradation due to the throughput of the binary shaped pupil mask.

Table 1. Parameters of SCI and the JWST coronagraphs

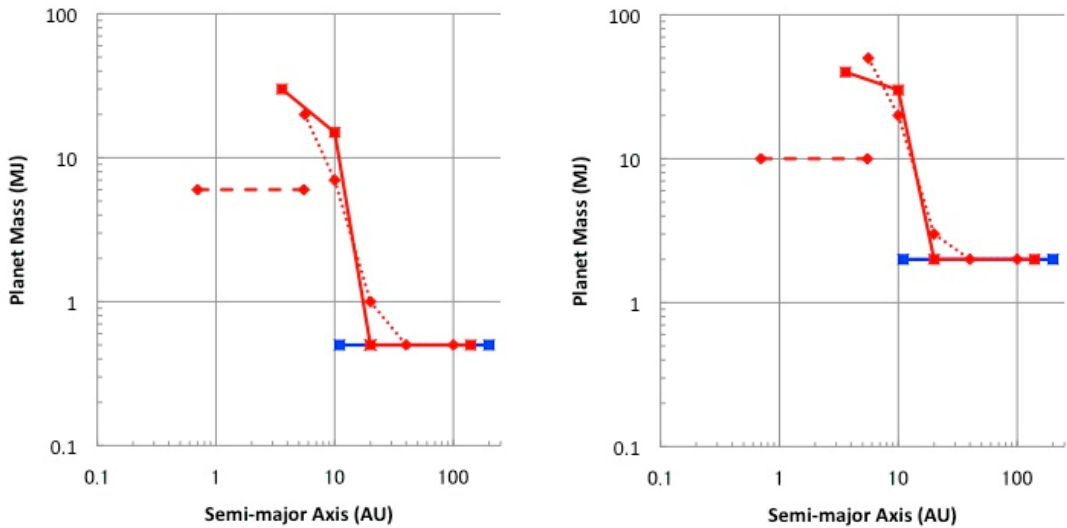

Figure 1. Expected limiting performances for planet detection around 1Gyr (left) and 5Gyr (right) old G stars with SCI (blue solid line), the FGS-TFI/NRM at 4.4 μm (red dashed line), the NIRCam coronagraph at 4.4 μm (red solid line), and the MIRI/FQPM at 11.4 μm (red dotted line), based on the COND03 model.

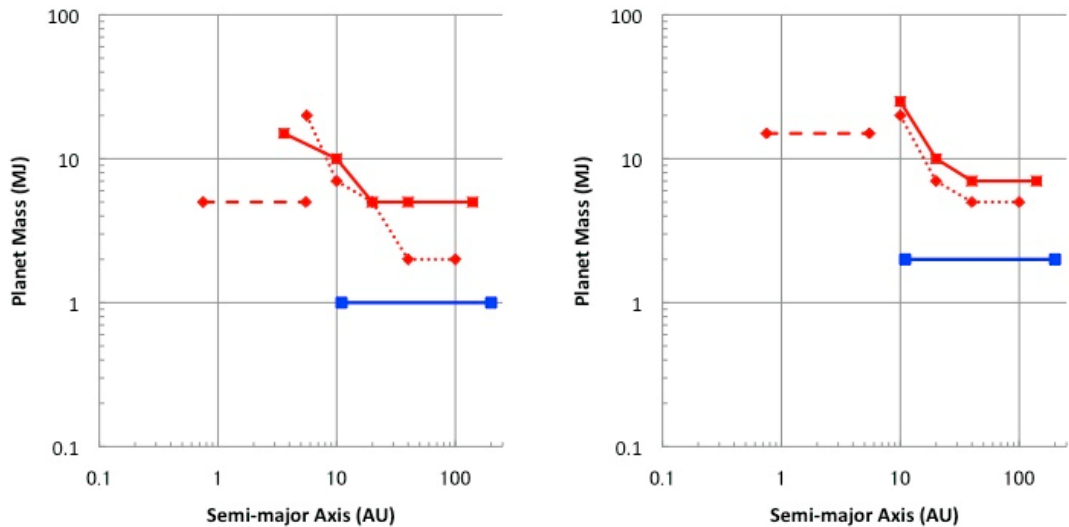

Figure 2. Expected limiting performances for planet detection around 1Gyr (left) and 5Gyr (right) old G stars with the SPICA and the JWST coronagraphs, based on the TLUSTY model. The lines are the same as those in Figure 1.

2.2. Characterization

Regarding the characterization of gas giants, SCI has several advantages over the JWST coronagraphs because the observing wavelengths and the spectral resolutions of the JWST coronagraphs are optimized only for planet imaging and thus its observing wavelengths and spectral resolutions are limited. On the other hand, SCI allows us to perform spectroscopic observations in the observing wavelength range from 3.5 to 27 μm with spectral resolutions of 20 and 200. Figure 3 shows a comparison of SCI's sensitivity for low spectroscopic observations (R=20) with the model spectra for ten Jupiter-mass planets at 100 Myr, 300 Myr, 1 Gyr, and 5 Gyr, calculated by Burrows et al. (2003). There are several absorption features in the mid-infrared: the methane features at 2.2, 3.3, and 7.8 μm, the ammonia features around 2.95 and 10.5 μm, and the water vapor features around 2.5 and 6 μm. As shown in Figure 3, SCI allows us to detect the water vapor absorption lines around 6 μm, the methane absorption line at 7.8 μm, and the ammonium absorption line at 10.5 μm for cold gas giants of less than 1 Gyr. Thus, SCI can fully characterize the very cool gas giants and accurately determine their temperatures.

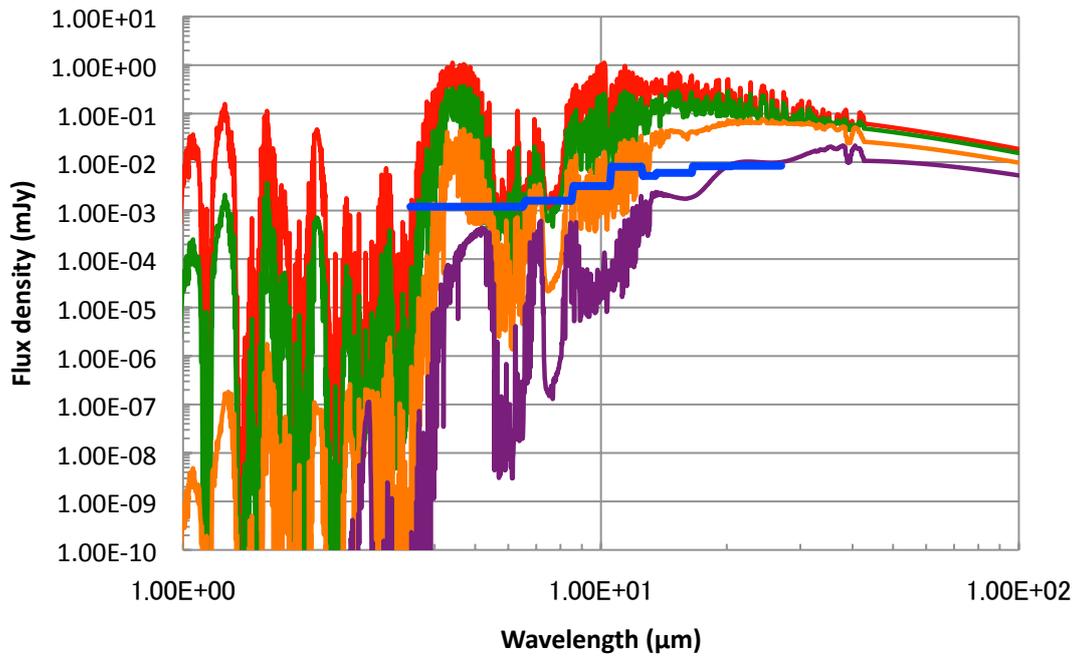

Figure 3. Comparison of the sensitivity of SCI's low spectroscopic observation R=20 (blue solid line) in one-hour integration time with the atmospheric model for $2M_J$ planets of 100 Myr (red solid line), 300 Myr (green solid line), 1 Gyr (orange solid line), and 5 Gyr (purple solid line) at 10 pc, based on the TLUSTY model of Burrows et al. (2003).

2.3. Detection of outer icy giants around early-type stars

In this section, we discuss planet detection around early-type stars. The reason why we focus on this is that planets orbiting early-type stars are warm due to the irradiation from their host stars even if these planets are beyond 10 AU. In addition to this, the planet's thermal emission relaxes the required contrast ratio of the planet to the host star at observing wavelengths longer than 10 μm. Therefore, focusing on the advantage of SCI's high dynamic range at observing wavelengths of longer than 10 μm, SCI can potentially detect outer icy giants or even super-Earths irradiated by early-type stars.

In order to evaluate SCI's detectability of icy gas giants or super-Earths around early-type stars, we take two systems, Altair (A7V) at 5.14 pc, Vega (A0V) at 7.76 pc, as examples. Altair and Vega are taken to be blackbody objects with effective temperatures of 6900 K and 9602 K and luminosities of 10.6 Lsun and 37 Lsun, respectively. The physical sizes of Altair and Vega are 1.6 and 2.26 times the solar radii, respectively. On the other hand, the target planets are taken to be blackbody objects with a radius range from 1.0 to 6.0 times the Earth radii. For simplicity, we assume that the effective temperature of the planet is characterized by the semi-major axis and the luminosity of host star from the radioactive equilibrium:

$T_p = 68K \left( \dfrac{a}{20.1 AU} \right)^{-1/2} \left( \dfrac{L}{1L_e} \right)^{1/4}$ , where $a$ is the semi-major axis of planet and

$L$ is the luminosity of host star. We adopt the average surface temperature of Uranus because Uranus does not have an additional internal thermal source.

Figure 4 shows SCI's limiting performances for planet detection around Altair and Vega. SCI allows us to detect the icy giants like Uranus around Altair and Neptune around Vega. On the other hand, extremely high contrast is required for direct detection of the icy giants in the near infrared range because the icy giants shine by reflection of light from their host stars. Near-infrared high contrast experiments, including JWST and the Extreme Large Telescopes (ELTs), cannot reach these icy giants because of the lack of contrast. Thus, SCI provides us with the first opportunity for discovering the outer icy giants although there are only four A-type stars within 10pc.

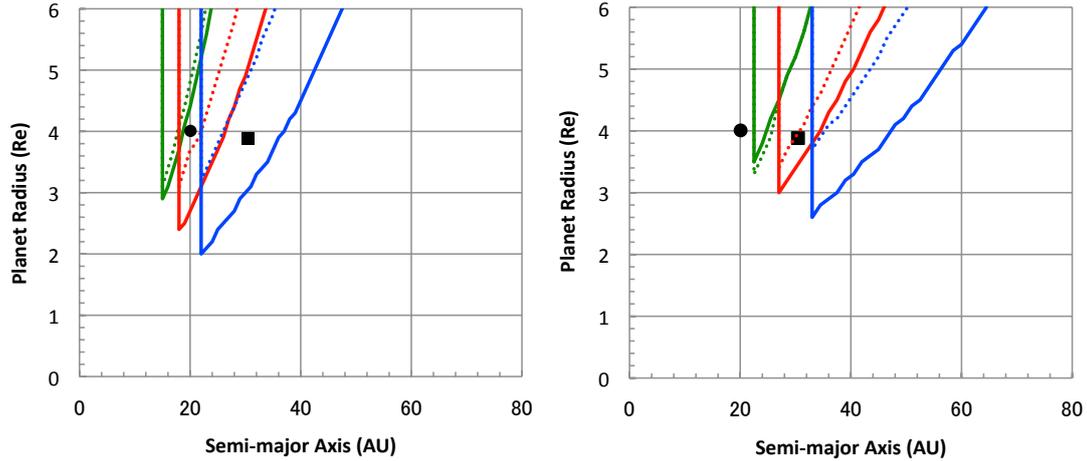

Figure 4. Limiting performances of SCI for planet detection around Altair (left) and Vega (right) in the wavelength ranges from 15 to 18 μm (green lines), 18 to 22 μm (red lines), and 22 to 27 μm (blue lines). The solid and the dotted lines show the contrast and the sensitivity limits, respectively. The black filled circles and squares represent Uranus and Neptune.

## 3. Approach to questions concerning the origin and evolution of planetary systems with SCI

As described in Section 1, only a few gas giants have been discovered so far by direct imaging. These gas giants are very massive (~10 Jupiter masses) and their orbits are far (~10 to a few tens AU) from the host stars. Such unexpected discoveries raise new questions about the origin of these gas giants. In Section 3, based on the derived limiting performance of SCI, we examine how to approach the questions concerning the formation and evolution of the outer gas giants from observations made with SCI.

### 3.1. Semi-major axis

There are two representative planet formation models, the Core-Accretion (CA) scenario (e.g., Safronov 1969; Goldreich & Ward 1973; Hayashi et al. 1985; Pollack et al. 1996) and the Disk Instability (DI) scenario (e.g., Kuiper 1951; Cameron 1978). According to the core-accretion model, a heavy element core is formed by the accretion of planetesimals. As the core grows, its ability to accrete gas from the surrounding disk increases.

When the core is sufficiently massive, rapid gas accretion occurs onto the core, and a gas giant is formed. On the other hand, according to the disk instability scenario, if a disk is sufficiently massive, the disk fragments into dense cores. Such clumps can contract to form giant gaseous protoplanets in several hundred years.

Here, in order to observationally constrain the formation mechanism of outer gas giants, we examine the conditions for gas giant formation in the two models. In this section, we investigate the limiting distances from the host stars for the two models. The gas giants are likely to be formed near their host stars through the core-accretion mechanism because a massive core cannot be fully formed in the outer region due to the low surface density of the solids. Dodson-Robinson et al. (2009) and Rafikov (2010) derived the limiting distance for the core-accretion model. Assuming that the critical core-mass is 10 times the mass of the Earth, the limiting distance is less than 20 AU for a medium-mass disk, with the Toomre Q-parameter = 8. Even if the core is formed in a very massive disk, with Q=1.5, the limiting distance is around 40 AU. On the other hand, disk fragmentation tends to occur in the outer disk, at more than 20-40 AU, because the high temperature of the inner disk provides high pressure against gravitational instability (e.g., Rafikov 2005; Boley & Durisen 2008; Stamatellos & Whitworth 2008; Cai et al. 2010). Therefore, the semi-major axis of the planet characterizes the two planetary formation scenarios (see Figure 5).

After the gas giants are formed, several mechanisms such as planet-planet scattering (e.g., Weidenschilling & Marzari 1996) and outward migration through planet-disk interaction, called Type-III migration (e.g., Lin, M.-K. & Papaloizou 2010), determine the evolution of the orbits of gas giants. These mechanisms can move the gas giants formed through the core-accretion mechanism into wide orbits. However, Peplinski et al. (2008) found that there is no possibility of moving the gas giants away to 60AU through numerical simulation of type-III migration. On the other hand, the gas giants scattered by planet-planet interactions are ejected from the planetary system with the exception that the orbit of the gas giants scattered is stabilized by planetesimals (e.g., Dodson-Robinson et al. 2009). Therefore, it is very difficult to move the gas giants formed through the core-accretion mechanism into the orbits wider than 60AU. Thus, even if the orbit of a gas giant has evolved, the two planetary formation mechanisms are still characterized by the semi-major axis (see Figure 5).

Here, focusing on the fact that SCI can detect planetary-mass objects in both the young and the old planetary systems shown in Figure 1, SCI will provide a great contribution to exoplanet study. This is because SCI can reveal both the "initial" and "final" distributions of the semi-major axes of the gas giants (see Figure 2). SCI allows us to observationally constrain not only the planetary formation model but also the planetary evolution mechanism.

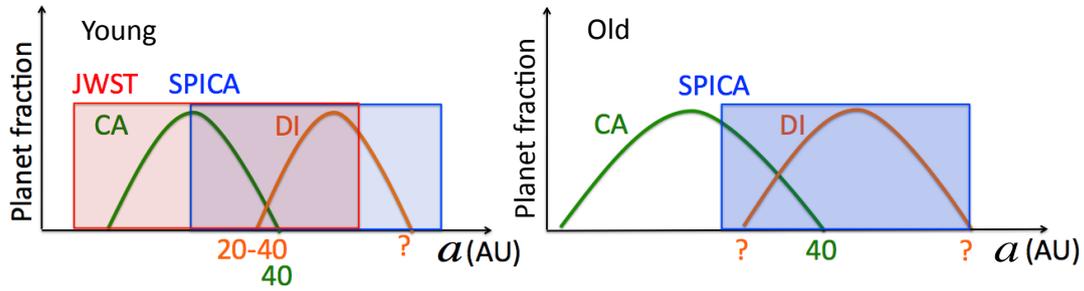

Figure 5. Comparisons of planet fractions expected from the core-accretion (red solid line) and the disk instability (green solid line) models with the detectable ranges of the planetary-mass objects by SCI (blue box) and JWST (red box) as a function of the semi-major axis for two epochs, just after the formation of planets (left) and after the evolution of them through planet-planet scattering and the outward migration due to planet-disk interaction (right). The reason why the detectable range of JWST is missing in the right panel is that the JWST coronagraphs are insensitive to planetary-mass objects in old planetary systems.

### 3.2. Icy giants

In Section 2.3, we found that direct imaging of the icy giants with SCI is possible. According to the standard scenario for the formation of the solar system (e.g., Hayashi et al. 1985), icy cores in the outer region cannot grow sufficiently through the accretion of planetesimals before gas in the disk depletes because the core-accretion rate is very small. As a result, the gas cannot adequately accrete onto the massive icy cores even if the icy cores are very massive. Ida & Lin (2004) theoretically showed that the icy gas giants as well as the terrestrial planets can be formed through the core-accretion mechanism. On the other hand, gravitational waves arising from the disk instability drive significant inward migration of the cores preventing the formation of massive cores including icy giants within the unstable disk (see Boss 1998). Therefore, once the disk is unstable, it is very difficult to form the icy giants because the disk instability occurs in the outer disk, where the icy giants can be formed, as discussed in Section 3.1. Thus, the presence of icy giants is proof that the planetary system is built through the core-accretion process. SCI will provide clues for the planet formation scenario but this issue cannot be statistically discussed from the point of view of the icy giants.

4. Conclusion

We derived and compared the limiting performances for planet detection of SCI and the JWST coronagraphs, based on the atmospheric evolution models provided by Baraffe et al. (2003) and Burrows et al. (2003). Based on the calculation of Baraffe et al. (2003), very cool objects down to an effective temperature T=100K can be detected by the NIRCam coronagraph, the MIRI/FQPM, and SCI for both young (1 Gyr) and old (5 Gyr) systems. In this case, the combination of the JWST coronagraphs is better than SCI for planet detection because its combination can detect the planetary mass objects both near and far from the host stars. On the other hand, according to the calculation of Burrows et al. (2003), a greater dynamic range is required for planet detection. As a result, SCI can detect even very cool objects orbiting old host stars by use of the advantage of SCI's much higher dynamic range at the observing wavelengths longer than 10 μm, while the JWST coronagraphs can detect massive planetary objects only far from their host stars due to the smaller dynamic range. Thus, SCI has an advantage in detecting planets around old host stars.

We also evaluated the capability of spectroscopic observations with SCI. Thanks to its wide observing wavelength range from 3.5 to 27 μm and high sensitivity, SCI can detect the water vapor feature around 6 μm, the methane feature at 7.8 μm, and the ammonia feature at 10.5 μm from the atmospheres of very cool objects. Therefore, SCI can fully characterize the gas giants and accurately determine their temperatures.

Furthermore, we predicted that the first direct imaging of icy giants around early-type stars will be made with SCI. Focusing on fact that the outer warm icy giants irradiated by early-type stars ease the required contrast ratio of the icy giant to the host star, we found that the outer icy giants like Uranus and Neptune can be detected around Altair and Vega, respectively. These searches with SCI will develop our understanding of the formation of outer icy giants.

Based on these evaluations, we discussed a new approach to answering questions concerning the origin and evolution of the outer planets. SCI will be able to fully observe both the initial and the final distributions of the gas giants in terms of the semi-major axis. These observations will lead us directly to understand when, where, and how the outer planets form and evolve. Thus, the SPICA coronagraph is complementary with the JWST coronagraphs and other exoplanet observations.

Acknowledgment
We are very grateful to I. Baraffe and A. Burrows for providing the atmospheric evolutional models. We greatly thank an anonymous referee whose careful reading improves this article. TM is financially supported from the Japan Society for the Promotion of Science (JSPS).